\documentclass[reprint,twocolumn,showpacs,preprintnumbers,amsmath,aps,prd,amsfonts,10pt]{revtex4-1}
\usepackage{eso-pic}
\usepackage{graphicx}
\usepackage{color}
\usepackage{type1cm}
\newsavebox{\longversionbox}

\begin{document}

\title{A Multicomponent Dark Matter in a Model with Mirror Symmetry with Additional Charged Scalars}

\author{Mirza Satriawan}
\affiliation{Department of Physics, Universitas Gadjah Mada, Bulaksumur BLS 21 Yogyakarta 55281, Indonesia}
\date{\today}
\begin{abstract}
A model with a mirror symmetry whose particles content consist of the ordinary SM particles (plus the right handed neutrinos) and their parity mirror partners, can provide a multicomponent dark matter consist of cold and warm dark matter components. I add to the original mirror model a singlet scalar and its mirror partner, whose quantum numbers are the same as the singlet right handed electron (and its mirror-partner). The new scalar can have a zero VEV, while its mirror partner VEV is non zero.  As consequences  mirror photon will obtain mass whose order is around the neutral mirror weak boson mass, rendering the mirror electromagnetic-like interaction similar like a mirror weak interaction.  There is a mixing among the ordinary neutrinos, mirror neutrinos, the singlet and the doublet mirror electrons.  As a result the mirror doublet electrons can have masses in the keV order, becoming the warm dark matter component of this model.  The cold dark matter component comes from the mirror nucleons that can have mass larger than the ordinary nucleons. The Big Bang Nucleosynthesys constraint can be avoided by a large entropy production in the ordinary sector due to a slow decay of mirror singlet electrons.  The temperature ratio of the two sectors is approximately proportional to the ratio of the VEV's of the ordinary and mirror Higgs, and this also will determine the cold-warm dark matter contribution to the cosmic energy density.
\end{abstract}
\maketitle

\section{Introduction}
Even though a warm dark matter model can solve the small scale structure problems (the missing satellite, the cups-core, and the too big to fail problems), it cannot at the same time fulfill the Lyman-$\alpha$ constraint (see for ex.~\cite{lyman, lyman2}).  This has led some people to consider a mixed dark matter model, i.e. a mixture of cold and warm dark matters, which can solve the small scale structure problem while at the same time still fulfilling the Lyman-$\alpha$ constraint (see for ex.~\cite{mixdark1,mixdark2,mixdark3}).  We also know that the dark matter energy density is of the same order as the baryonic energy density, $\Omega_{DM} \approx 5 \ \Omega_B$~\cite{planck}, and we know that the value of $\Omega_B$ is the outcome of several physical phenomena, i.e. the baryogenesis/leptogenesis, and several Standard Model (SM) physics including hadronization, the nucleon masses and the stability of proton.  All of the above suggest that the dark matter sector maybe a rich multi-component sector of cold and warm dark matter, that has some similarity to the SM, but with some differences to account for larger energy density and other constraints.   

Following this suggestion the mirror sector of the mirror model~\cite{mirrorfoot1, mirrorfoot2} seems to be a good candidate for this multi-component dark matter sector (For a review see for example~\cite{mirrorfoot3, mirrorfoot4}).  The mirror model particle content are the ordinary particle, i.e. the SM particle content (plus additional right handed singlet neutrinos) plus  their mirror partners.  The Lagrangian of the model is invariant under the gauge group SU(3)$_1$$\otimes$SU(3)$_2$$\otimes$SU(2)$_L$$\otimes$SU(2)$_R$$\otimes$U(1)$_{Y}$$\otimes$U(1)$_{X}$ and the Z$_2$-mirror symmetry that transform a left (or right) chiral ordinary particle into its right (or left) chiral mirror partner and vice versa.  In its original version, the model has a mirror partner of the Higgs scalar, with the same vacuum expectation value (VEV).  As consequences, the mass spectra of the mirror particles is the same as in the ordinary particles.  Correspondingly, the mirror neutrino and the ordinary neutrino are mixed maximally, contradicting the three SM-neutrino oscillation scheme.  Besides this, the long range electromagnetic-like interaction in the mirror sector is in tension with the nature of the dark matter self interaction inferred from the bullet cluster observation~\cite{bullet}.  Moreover, to escape the big bang nucleosynthesis (BBN) constraint, the mirror sector has to be colder than the ordinary sector, and in the original mirror model, this temperature difference can only come from a process in the inflation and reheating era~\cite{bereziani}.  

Some people have proposed modifications of this mirror model by setting different VEVs for the Higgs and its mirror partner and giving some mass to the mirror photon~\cite{anh,anh2}.  The mass that is given to the mirror photon usually small, less than the mass of the mirror electrons, so that mirror electrons will still pair annihilate into mirror photons, thus their density will not over-close the universe.   

In this paper I propose a modified mirror model, also by making the mirror photon massive but with a large mass larger than the electroweak scale.  This is done by adding into the model, a scalar (and its mirror partner)  whose gauge quantum numbers are the same as the right handed electron (and its mirror partner).  When the mirror partner of this scalar gain a non zero VEV, the gauge field of U(1)$_X$ will gain mass, thus in the end giving a large mass to the mirror photon.  A non zero VEV of this new scalar will break the mirror symmetry of the scalar potential such that the mirror Higgs can have a different non zero VEV than the Higgs scalar.   It turns out also that the new scalar, being interacting with some mirror fermions, can lead to some mirror fermions having masses in the keV order, thus providing the candidate for keV-warm dark matter. The BBN constraint can be solved either by using the physics in the reheating era or using a slow decay of the massive m-singlet electron, that decay dominantly into the ordinary sector than the mirror sector, thus increasing the ordinary sector temperature relative to the mirror sector.  

It is reasonable to assume that this modified mirror model is just a low energy version of some Grand Unified Theory with a non abelian gauge group, and therefore there is no mixing between the two U(1)'s gauge bosons.  The complete fermion and scalar particles with its mirror partner of the modified mirror model are given in Table \ref{tabelmassa1}.   In the following the prefix o- and m- refer to ordinary and mirror respectively.
\begin{table}
	\centering
	\caption{Irreducible representation (irreps) and quantum numbers assignment for the scalar and spinor particles with respect to the mirror gauge group.  Note: the last two lines are the scalars.}
\begin{tabular}{c c c c c}\hline\hline
o-particles & Irreps & & m-particles & Irreps \\ \hline\hline
 & & & & \\
$L_L \equiv \left( \begin{array}{l}
\nu \\
e
\end{array}\right)$ & (\textbf{1},\textbf{1},\textbf{2},\textbf{1},-1,0) & &
$L_R \equiv \left( \begin{array}{l}
N \\
E
\end{array}\right)$ &(\textbf{1},\textbf{1},\textbf{1},\textbf{2},0,-1) \\ 
$\nu_R$ &(\textbf{1},\textbf{1},\textbf{1},\textbf{1},0,0)& & $N_L$ & (\textbf{1},\textbf{1},\textbf{1},\textbf{1},0,0) \\ 
 $e_R$ & (\textbf{1},\textbf{1},\textbf{1},\textbf{1},-2,0)& &$E_L$ & (\textbf{1},\textbf{1},\textbf{1},\textbf{1},0,-2) \\ 
$Q_L \equiv \left( \begin{array}{c}
u \\
d
\end{array}\right)$ &(\textbf{3},\textbf{1},\textbf{2},\textbf{1},$\frac{1}{3}$,0) & &
$Q_R = \left( \begin{array}{c}
U \\
D
\end{array}\right)$ &(\textbf{1},\textbf{3},\textbf{1},\textbf{2},0,$\frac{1}{3}$)\\
$u_R$ & (\textbf{3},\textbf{1},\textbf{1},\textbf{1},$\frac{4}{3}$,0)& & $U_L$ & (\textbf{1},\textbf{3},\textbf{1},\textbf{1},0,$\frac{4}{3}$)\\ 
$d_R$ & (\textbf{3},\textbf{1},\textbf{1},\textbf{1},$\frac{-2}{3}$,0)& &
  $D_L$ & (\textbf{1},\textbf{3},\textbf{1},\textbf{1},0,$\frac{-2}{3}$) \\
			& & & & \\\hline 
		& & & & \\
	$\chi_L = \left( \begin{array}{c}
\chi_\nu \\
\chi_e
\end{array}\right)$ &(\textbf{1},\textbf{1},\textbf{2},\textbf{1},-1,0)& &
$\chi_R = \left( \begin{array}{c}
\chi_N \\
\chi_E
\end{array}\right)$ &(\textbf{1},\textbf{1},\textbf{1},\textbf{2},0,-1)\\
 $\phi_e$ & (\textbf{1},\textbf{1},\textbf{1},\textbf{1},-2,0)& &$\phi_E$ & (\textbf{1},\textbf{1},\textbf{1},\textbf{1},0,-2) \\ 
	& & & & \\\hline\hline
\end{tabular}
\label{tabelmassa1}
\end{table}
\section{The Scalar Potential}
The most general scalar potential which is invariant under the gauge and Z$_2$-mirror transformation is
\begin{eqnarray}
&V_{\mathcal{H}}& = -\mu_1^2 \left( \left | \chi_L \right| ^2 + \left | \chi_R \right| ^2 \right)  -\mu_2^2 \left(\left | \phi_e \right| ^2 + \left | \phi_E \right| ^2 \right) \nonumber\\&+& \lambda_1 \left( \left | \chi_L \right| ^4 + \left | \chi_R \right| ^4 \right) + \lambda_2 \left( \left | \phi_e \right| ^4 + \left | \phi_E \right| ^4 \right) \nonumber\\&+&\alpha_1  \left | \chi_L \right| ^2 \left | \chi_R \right| ^2 +  \alpha_3 \left( \left | \chi_L \right| ^2 \left | \phi_E \right| ^2+ \left | \chi_R \right| ^2 \left | \phi_e \right| ^2 \right) \nonumber\\ &+& \alpha_2 \left| \phi_E \right| ^2 \left | \phi_e \right| ^2 + \alpha_4 \left( \left | \chi_L \right| ^2 \left | \phi_e \right| ^2+ \left | \chi_R \right| ^2 \left | \phi_E \right| ^2 \right),  \label{higgspotensial}
\end{eqnarray}
where the $\mu_i$'s, $\lambda_i$'s and $\alpha_i$'s are parameters of the potential.  The $\lambda_i$'s and the $\alpha_i$'s have to be positive in order for the potential to be bounded below.  The potential parameters can have values such that at higher energy there is a spontaneous symmetry breaking that left $\phi_E$ to have a non zero VEV, $\langle \phi_E \rangle = v_E$, while $\phi_e$ remains with a zero VEV.  

At a lower energy another spontaneous symmetry breaking makes the $\chi_R$ and $\chi_L$ acquired non zero VEVs.  To fulfill local minimum value of the scalar potential, the VEVs are related as follows
\begin{equation}
\langle \chi_R \rangle^2 \equiv v_R^2 = \frac{2\mu_1^2}{2\lambda_1+\alpha_1} + \frac{\alpha_1\alpha_3 - 2\lambda_1\alpha_4}{4\lambda_1^2 - \alpha_1^2} v_E^2,
\end{equation}
\begin{equation}
\langle \chi_L \rangle^2 \equiv v_L^2 = \frac{2\mu_1^2}{2\lambda_1+\alpha_1} + \frac{\alpha_1\alpha_4 - 2\lambda_1\alpha_3}{4\lambda_1^2 - \alpha_1^2} v_E^2,
\end{equation}
from which we have $v_R^2 - v_L^2 = (\alpha_3-\alpha_4)v_E^2/(2\lambda_1-\alpha_1)$.  Therefore if $\alpha_3,\alpha_4 << 1$ we can have $v_E >> v_R, v_L$ and $v_R \neq v_L$. 

After $\chi_R$ and $\chi_L$ gain VEVs, the $\phi_e$ can acquired mass, given at the classical level by
\begin{equation}
m^2_{\phi_e} = \frac{1}{2}\left(\alpha_3 v_R^2 + \alpha_4 v_L^2 \right)
\end{equation}
The other three scalars will form the following mixing mass term $\mathbf{H}^T \mathbf{M}_h \mathbf{H}$ where $\mathbf{H}\equiv(h_E,h_R,h_L)^T$ are the scalar field excitation above their VEV's, and 
\begin{equation}\label{massmatrix}
 \mathbf{M}_h =v_R^2\left( \begin {array}{ccc} \lambda_2/ \eta^{2}& \alpha_4 /2\eta & \alpha_3 \xi/2\eta \\ \noalign{\medskip}
\alpha_4/2\eta &\lambda_1 & \alpha_1 \xi /2\\ \noalign{\medskip} \alpha_3 \xi/2\eta &\alpha_1 \xi/2 & \lambda_1 \xi^2 \end {array} \right),
\end{equation}
where $\eta \equiv v_R/v_E$ and $\xi \equiv v_L/v_R$.  The current result of the 125 GeV Higgs measurement, does not allow us to put a strict constraint on the mixing of the Higgs and other scalars.  Nevertheless we can assume that there is no large mixing between the Higgs and other scalars, and that $v_R > v_L$.   Assuming that the parameters $\lambda$'s at the maximum are of order unity and $\alpha$'s are small, the above mass matrix should have eigenvalues $m_{h_1}>>m_{h_2}>m_{h_3}$ for corresponding mass basis $h_1, h_2,$ and $h_3$.  Moreover, $h_E$ is dominated by $h_1$, $h_R$ is dominated by $h_2$, and $h_L$ is dominated by $h_3$, with small mixing between three of them.  The $h_L$ is our o-Higgs, while $h_R$ is the m-Higgs.

\section{The Gauge Sector}
The mass of the gauge bosons comes from the usual gauge-scalar field interaction in the Lagrangian, which after symmetry breaking is written as
\begin{eqnarray}
\cal{L}&\supset&\frac{1}{4}v_L^2g^2 W_{L\mu} ^+ W^{-\mu}_L+\frac{1}{4}v_R^2g^2 W_{R\mu} ^{+} W^{-\mu}_R 
+\frac{1}{8}\mathcal{W}_L^T \mathbb{M}_L \mathcal{W}_L\nonumber\\&+&\frac{1}{8}\mathcal{W}_R^T \mathbb{M}_R \mathcal{W}_R \label{matriks}
\end{eqnarray}
where $\mathcal{W}_L =(W_{L\mu}^3 , B_{Y\mu})^T $, $\mathcal{W}_R =(W_{R\mu}^3, B_{X\mu} )^T$ and  
\begin{equation}\label{WRL}
W_{R\mu } ^{\pm} = \frac{(W^{1}_{R\mu} \mp iW^{2}_{R\mu})}{\sqrt{2}};\quad W_{L\mu} ^\pm =\frac{(W^{1}_{L\mu } \mp iW^{2}_{L\mu})}{\sqrt{2}} 
\end{equation}
with $\mathbb{M}_L$ is the same as in SM, while $\mathbb{M}_R$ is given by
\begin{eqnarray}\label{massright}
\left( \begin {array}{cc}{g}^{2}{v_R}^{2} &-gg'{v_R}^{2}
\\\noalign{\medskip}-gg'{v_R}^{2} &{g'}^{2}({v_R}^{2}+{4v_E}^{2}) 
\end {array} \right). 
\end{eqnarray}
The mass matrix above can be diagonalized using $\mathcal{W}_R = \mathbf{S}_R \mathcal{W'}_R$, where $\mathcal{W'}_R=(Z_R^\mu,D^\mu)^T$ is the mass basis, and
\begin{eqnarray}\label{matrikscampuran}
\mathbf{S}_R=\left(\begin {array}{cc} \frac{x\eta^2}{c_+} &\frac{x\eta^2}{c_-}\\\noalign{\medskip} -\frac{a+b}{c_+}   & \frac{b-a}{c_-}   \end {array} \right),  
\end{eqnarray}
with $x = g'/g\equiv \tan \theta_W$, and
\begin{eqnarray}\label{param}
a&=&\frac{1}{2}\left((x^2 -1)\eta^2+ 4x^2\right);\ 
b=\sqrt{a^2 + \eta^4 x^2};\nonumber\\ c_+ &=& \sqrt{2b(b+a)};\quad c_- =\sqrt{2b(b-a)}.  
\end{eqnarray}
The resulting gauge boson mass eigenvalues in the m-sector are
\begin{eqnarray}
m_{W_R} &=& \frac{gv_R}{2};\quad m_{Z_R}=\frac{gv_R}{2} \sqrt{1+\frac{\eta^2 x^2}{b-a}}\label{massawzr}\\
m_D  &=& \frac{gv_R}{2}\sqrt{1-\frac{\eta^2 x^2}{b+a}}.
\end{eqnarray}
The $D^\mu$ is the gauge field of m-electromagnetic interaction, i.e. the m-photon.  For the case when $\eta <<1$ $(v_E>>v_R)$, we have $m_D \rightarrow gv_R/2$ and $m_{Z_R} \rightarrow g'v_E$, thus at the lower energy the m-electromagnetic interaction can becomes as weak as the m-weak interaction, while the m-weak interaction is $\xi^4$ times weaker than the o-weak interaction.  

In term of mass basis, the current interaction between the m-gauge fields and the m-fermions can be written as follows
\begin{eqnarray}\label{current}
&-&i \frac{g c_+}{2b} \left(J_{\mu R}^3+\frac{x c_-}{ c_+}  \frac{J_\mu^X}{2}\right)D^\mu \nonumber\\ &-&i \frac{g c_-}{b-a} \left(J_{\mu R}^3-\frac{a+b}{2b} \left( J_{\mu R}^3 +\frac{x c_-}{ c_+}\frac{J_\mu^X}{2}\right) \right)Z_R^\mu,\nonumber\\
\end{eqnarray}
where $J_\mu$'s are the corresponding current.  We can identify from the term that coupled to $D^\mu$, the unit charge of the m-electromagnetic interaction, i.e. $gc_+/2b$, and the m-electromagnetic charge operator, given by
\begin{equation}
Q_{D} = T_{3R} + \frac{x c_-}{c_+} \frac{X}{2}.
\end{equation}
Except for $N_L$, all the other m-fermions have fractional m-electromagnetic charge.  When $\eta <<1$ the unit charge $gc_+/2b \rightarrow g$, $Q_D \rightarrow T_{3R}$ and the m-electromagnetic interaction will be similar to m-weak interaction.  I will assume the case $\eta << 1$ for the following.
	
\section{Yukawa Interaction}
The most general Yukawa interaction invariant under the gauge and Z$_2$-mirror symmetry, is (suppressing the generation index)
\begin{widetext}
\begin{eqnarray}
\mathcal{L} &&\supset - G_e \left( \bar{L}_L \chi_L ^c e_R + \bar{L}_R \chi_R ^c E_L  \right ) - G_\nu \left( \bar{L}_L \chi_L \nu_R  + \bar{L}_R \chi_R N_L  \right)- G_d \left( \bar{Q}_L \chi_L^c  d_R + \bar{Q}_R \chi_R^c  D_L \right ) - G_u \left( \bar{Q}_L \chi_L  u_R + \bar{Q}_R \chi_R  U_L \right )\nonumber\\&&- G_{\nu e} \left( \bar{e}_R \phi_e N_L  + \bar{E}_L \phi_E \nu_R \right ) - G_{\nu e}' \left( \bar{e}_R \phi_e \nu_R^c  - \bar{E}_L \phi_E N_L^c \right ) - G_\nu'\left(\bar{L}_L \chi_L N_L^c  - \bar{L}_R \chi_R \nu_R^c \right)\nonumber\\&&- M_m(\bar{\nu}_R^c\nu_R - \bar{N}_L^cN_L)- M_d \bar{N_L}\nu_R   + h.c.
\end{eqnarray}
\end{widetext}
the couplings $G$'s and the $M$'s above are three by three matrices to account for the three generations.  The fermions will get their masses when the scalars get their VEVs.  The m-quarks will have masses $\xi^{-1}$ times the mass of o-quarks, and the small mixing between their generations can be accounted in the elements of $G_u$ and $G_d$.  The different generation of o-charged leptons are not mixing and this should be reflected in $G_e$ being a diagonal matrix.    While the o-neutrinos, the m-neutrinos and the m-charged leptons are mixing even in one generation.   Using a Majorana basis $\psi^T_\alpha\equiv(\nu, N,  E,  E',  \nu', N')^T\equiv(\nu_L+\nu_L^c,  N_R+N_R^c, E_R+E_R^c, E_L+E_L^c, \nu_R+\nu_R^c, N_L+N_L^c)^T$, the mixing mass term can be written as $\frac{1}{2}\bar{\psi}\mathcal{M}\psi$ with
\begin{eqnarray}\label{campuranmassa}
\mathcal{M}=\left(\begin {array}{cccccc}  0 & 0 & 0 & 0 & M_{\nu L} & M_{\nu L}'  \\\noalign{\medskip}0 & 0 & 0 & 0 & - M_{\nu R}' & M_{\nu R}   \\\noalign{\medskip} 0 & 0 & 0 & M_e & 0 & 0 \\\noalign{\medskip} 0 & 0 & M_e^T & 0 & M_E & -M_E'  \\\noalign{\medskip}  M_{\nu L}^T  & - M_{\nu R}'^T & 0 & M_E^T & \widehat{M}_m & {M}_d \\\noalign{\medskip} M_{\nu L}'^T & M_{\nu R}^T & 0  & - M_E'^T & {M}_d^T& -\widehat{M}_m  \end {array} \right) \nonumber\\
\end{eqnarray} 
is a $6\times 6$ partitioned matrix whose entries are the following $3\times 3$ sub matrices: $M_e = v_R G_e$, $M_{\nu L}=v_L G_\nu$, $M_{\nu R}=v_R G_\nu$, $M_{\nu L}'=v_L G_\nu'$, $M_{\nu R}'=v_R G_\nu'$, $M_E= v_E G_{\nu e}$, $M_E'=v_E G_{\nu e}'$, $\widehat{M}_m = 2M_m$.  The exact eigenvalues and eigenvectors for this mass matrix can be obtained numerically.  But with some natural assumptions about the order of the sub matrices, the $\mathcal{M}$ can be diagonalized approximately.  Lets assume the following hierarchy, 
$M_m \approx M_d >> M_E\approx M_E' >> M_e\approx M_{\nu R}\approx M_{\nu R}'>> M_{\nu L}\approx M_{\nu L}'$.  This assumption naturally comes from the previous assumption above about the value of $\eta$ and $\xi$; while $M_m$ and $M_d$, being unrestricted by the mirror gauge symmetry, should have the largest order.  Due to this assumption the above mass matrix can be block diagonalized approximately using seesaw mechanism method~\cite{seesaw,seesaw2,seesaw3,seesaw4}.  

First, denotes the $\mathcal{M}$ as 
\begin{equation}
\mathcal{M} = \left(\begin{array}{cc} A & B \\\noalign{\medskip} B^{T} & D \end{array}\right)
\end{equation}
where
\begin{eqnarray}
A &=& \left( 0 \right),\quad B = \left( \begin{array}{cc} 0 & \mathcal{M}_\nu \end{array}\right),\nonumber\\
D &=& \left( \begin{array}{cc} \mathcal{M}_e & \mathcal{M}_E \\\noalign{\medskip}  \mathcal{M}_E^T & \mathcal{M}_m\end{array}\right)
\end{eqnarray}
with the entries are the following block matrices
\begin{eqnarray}
\mathcal{M}_\nu &=& \left( \begin{array}{cc}  M_{\nu L} & M_{\nu L}' \\\noalign{\medskip}  - M_{\nu R}' & M_{\nu R} \end{array}\right),\quad
\mathcal{M}_e = \left( \begin{array}{cc} 0 & M_e \\\noalign{\medskip}  M_e^T & 0 \end{array}\right) \nonumber\\
\mathcal{M}_E &=& \left( \begin{array}{cc} 0 & 0 \\\noalign{\medskip}  M_E & -M_E' \end{array}\right),\quad
\mathcal{M}_m = \left( \begin{array}{cc}  \widehat{M}_m & M_d \\\noalign{\medskip}  M_d^T & -\widehat{M}_m \end{array}\right). 
\end{eqnarray}
Since the entries of $D$ is greater than that of $B$, we can use the seesaw mechanism to get
\begin{equation}\label{Mpertama}
\mathcal{M} \approx U \left(\begin{array}{cc} \mathcal{M}_\nu \mathcal{M}_m^{-1}\mathcal{M}_\nu^T & 0 \\\noalign{\medskip} 
0 & D \end{array}\right)U^T
\end{equation}  
where
\begin{equation}\label{Upertama}
U \approx \left(\begin{array}{cc} I & BD^{-1T} \\\noalign{\medskip} -D^{-1}B^T & I \end{array} \right)
\end{equation}
The content of the upper left block in \eqref{Mpertama} is 
\begin{equation}
\mathcal{M}_\nu \mathcal{M}_m^{-1}\mathcal{M}_\nu^T = \left( \begin{array}{cc} A_a & B_a \\\noalign{\medskip} B_a^T & D_a  \end{array} \right)
\end{equation}
where
\begin{eqnarray}\label{ABD}
A_a &=& M_{\nu L}' \widehat{M}_m^{-1} M^{'T}_{\nu L}- M_{\nu L} \widehat{M}_m^{-1} M_{\nu L}^T  \nonumber\\
B_a &=& M_{\nu L} \widehat{M}_m^{-1} M^{'T}_{\nu R} + M_{\nu L}' \widehat{M}_m^{-1} M_{\nu R}^T \nonumber\\
D_a &=& M_{\nu R} \widehat{M}_m^{-1} M_{\nu R}^T - M_{\nu R}' \widehat{M}_m^{-1} M^{'T}_{\nu R}.
\end{eqnarray}
Because $v_R >> v_L$ then $D_a>> B_a>> A_a$, thus we can use seesaw mechanism to write
\begin{equation}
\mathcal{M}_\nu \mathcal{M}_m^{-1}\mathcal{M}_\nu^T \approx U_a\left( \begin{array}{cc} A_a-B_aD_a^{-1}B_a^T & 0 \\\noalign{\medskip} 0 & D_a \end{array} \right)U_a^T
\end{equation}
where
\begin{equation}
U_a\approx \left( \begin{array}{cc} I & B_aD_a^{-1 T} \\\noalign{\medskip} - D_a^{-1}B_a^T & I \end{array}\right). 
\end{equation}

Next, consider the matrix $D$ in \eqref{Mpertama}, since the entries of $\mathcal{M}_e$ and $\mathcal{M}_E$ are very small compared to the entries of $\mathcal{M}_m$, we can use seesaw mechanism to obtain 
\begin{equation}\label{Dblock}
D \approx U' \left( \begin{array}{cc} \mathcal{M}_e - \mathcal{M}_E\mathcal{M}_m^{-1}\mathcal{M}_E^T & 0  \\\noalign{\medskip} 
0 & \mathcal{M}_m \end{array}\right) U'^T
\end{equation}
where
\begin{equation}
U'\approx \left( \begin{array}{cc} I & \mathcal{M}_E\mathcal{M}_m^{-1 T} \\\noalign{\medskip} - \mathcal{M}_m^{-1}\mathcal{M}_E^T & I \end{array}\right). 
\end{equation}
Now, consider the upper left block matrix in \eqref{Dblock},
\begin{equation}
\mathcal{M}_e - \mathcal{M}_E\mathcal{M}_m^{-1}\mathcal{M}_E^T
= \left(\begin{array}{cc} 0 & M_e \\\noalign{\medskip} M_e^T & D_b\end{array}\right),
\end{equation}
where $D_b = M_E' \widehat{M}_m^{-1}M^{'T}_E-M_E \widehat{M}_m^{-1}M_E^T$.  If we assume that the entries of $M_e$ is very small compare to $D_b$, we can use seesaw mechanism again, to write
\begin{equation}
D_a\approx U_b \left( \begin{array}{cc} M_e D_b^{-1} M_e^T & 0 \\\noalign{\medskip}  0 & D_b \end{array}\right) U_b^T,
\end{equation}
where 
\begin{equation}
U_b\approx \left( \begin{array}{cc} I & M_e D_b^{-1 T} \\\noalign{\medskip} - D_b^{-1}M_e^T & I \end{array}\right). 
\end{equation}
Certainly there is no specific reason why $M_e$ has to be very small compare to $D_b$, besides for simplicity and the applicability of seesaw mechanism.

Lastly, consider the lower right block matrix of \eqref{Dblock}, i.e $\mathcal{M}_m$.  We have assume that $M_m\approx M_d$, but for simplicity lets assume also that $\widehat{M}_m > M_d$, then we can write
\begin{eqnarray}
&&\mathcal{M}_m \approx\nonumber\\
&&U_c\left( \begin{array}{cc} \widehat{M}_m+M_d \widehat{M}_m^{-1}M_d^T  & 0 \\\noalign{\medskip}  0 & -(\widehat{M}_m+M_d \widehat{M}_m^{-1}M_d^T) \end{array}\right)U_c^T,\nonumber\\
\end{eqnarray}
for some matrix $U_c$ given by
\begin{equation}
U_c\approx \left( \begin{array}{cc} I & -M_d\widehat{M}_m^{-1} \\\noalign{\medskip} \widehat{M}_m^{-1}M_d^T & I \end{array}\right). 
\end{equation}
Collecting everything together and using the mass basis $\hat\psi_i\equiv(n_1, n_2, n_3, n_4, n_5, n_6)^T$ where each $n_i$'s is a three component vector, we can write the mixing mass term as $\frac{1}{2}\bar \psi \mathcal{M}\psi = \frac{1}{2}\bar{\hat\psi}\mathcal{M}_\delta\hat{\psi}$.  The $\mathcal{M}_\delta$ is a $6\times 6$ diagonal partitioned matrix whose diagonal block are the following $3\times 3$ matrices
\begin{eqnarray}\label{massneut}
\mathcal{M}_1 &=& V_1^T(A_a-B_aD_a^{-1}B_a^T)V_1,\quad \mathcal{M}_2 = V_2^TD_aV_2,\nonumber\\
\mathcal{M}_3 &=& V_3^TM_e D_b^{-1} M_e^TV_3 ,\quad \mathcal{M}_4 = V_4^TD_bV_4 ,\nonumber\\
\mathcal{M}_5 &=& V_5^T(\widehat{M}_m+M_d \widehat{M}_m^{-1}M_d^T)V_5 ,\nonumber\\
\mathcal{M}_6 &=& -V_6^T(\widehat{M}_m+M_d \widehat{M}_m^{-1}M_d^T)V_6,
\end{eqnarray}
with $V_i$'s are the matrices that will diagonalize the three generations mixing in each block. The mass matrices in \eqref{massneut} have been labeled in the order of increasing mass order.   The flavor basis $\psi$ is related to the mass basis through $\psi_\alpha = \mathcal{U}_{\alpha i}\hat{\psi_i}$, with $\mathcal{U}_{\alpha i}$ are the entries of the following matrix
\begin{equation}\label{neutrinomix}
\mathcal{U} \approx  U \left(\begin{array}{ccc} U_a & 0 & 0 \\\noalign{\medskip} 0 & U_b & U_b \mathcal{M}_E\mathcal{M}_m^{-1 T} \\\noalign{\medskip} 
0 & -U_c\mathcal{M}_m^{-1}\mathcal{M}_E^T & U_c\end{array} \right) \mathcal{V}
\end{equation}
with $U$ is given in \eqref{Upertama}.  While $\mathcal{V}$ is a $6\times 6$ block diagonal matrix whose block are $V_i$'s ($i=1,\dots, 6)$. For first order approximation, $U$ can be taken as a unit diagonal matrix, and thus
\begin{widetext}
\begin{equation}\label{neutrinomix}
\mathcal{U} \approx  \left(\begin{array}{cccccc} I & B_aD_a^{-1 T} & 0 & 0 & 0 & 0 \\\noalign{\medskip} 
- D_a^{-1}B_a^T & I & 0 & 0 & 0 & 0 \\\noalign{\medskip}
0 & 0 & I & M_e D_b^{-1 T} & M_e M_E^{-1}  & M_e M_E^{-1} \\\noalign{\medskip} 
0 & 0 & - D_b^{-1}M_e^T & I & M_E\widehat{M}_m^{-1} & M'_E\widehat{M}_m^{-1} \\\noalign{\medskip} 
0 & 0 & 0 & -\widehat{M}_m^{-1}M_E^T  &  I & -M_d\widehat{M}_m^{-1} \\\noalign{\medskip} 
0 & 0 & 0 & -\widehat{M}_m^{-1}M_E'^T &  \widehat{M}_m^{-1}M_d^T & I\end{array} \right) \mathcal{V}
\end{equation}
\end{widetext}
From the entries of $\mathcal{U}$ in \eqref{neutrinomix}, we can see that $\nu$ is dominated by $n_1$, and $N$  is dominated by $n_2$, and there is mixing between $\nu$ and $N$, whose probabilities are determined by the entries of $B_aD_a^{-1T}$.  Assuming $G_\nu\approx G_\nu'$ then the mixing probability between o- and m-doublet neutrinos are proportional to $\xi^2$.  The limit on the sterile neutrinos-neutrinos mixing from several collaborations~\cite{adamson1,adamson2,icecube,icecubedeep} give upper limit on $|U_{\mu 4}|^2 $ and $|U_{\tau 4}|^2$ that can be use to give the order of magnitude value of $\xi$, i.e. $\xi \approx 10^{-1}$ or less.  Lets assume this value for the following.    

The $E$ is dominated by $n_3$,  and $E'$ is dominated by $n_4$, and there is mixing between $E$ and $E'$ whose probabilities are determined by the entries of $M_eD_b^{-1T}$.  The $E'$ has a moderate mixing with $\nu'$ and $N'$, whose probabilities are determined by the entries of $M_E\widehat{M}_m^{-1}$.  Because $M_e<< D_b$ then $E$ only have smaller mixing with $\nu'$ and $N'$, with the probabilities are determined approximately by the entries of $M_e M_E^{-1}$.     Lastly the $\nu'$ and $N'$ are dominated by the mixture of $n_5$ and $n_6$.    

The matrix $\mathcal{M}_1$ is the usual neutrino mass matrix, with $V_1$ as the PMNS mixing matrix~\cite{pmns1, pmns2}.   The mass of m-doublet neutrino is determined by $\mathcal{M}_2$ in \eqref{massneut}, then using \eqref{ABD} the mass of m-doublet neutrinos should be $\xi^{-2}$ times the mass of o-doublet neutrinos.  Assuming normal neutrino hierarchy, the mass of the largest o-doublet neutrinos is around $10^{-1}$ eV, while the lightest o-doublet neutrinos should be below $10^{-2}$ eV.   Thus, for $\xi \approx 10^{-1}$ the largest mass of m-doublet neutrinos should be around $10$ eV, while the lightest m-doublet neutrinos should be below $1$ eV.  The $\mathcal{M}_2$ should be maximally mixed with $V_2 \approx V_1$, and at current cosmic temperature, the remaining m-doublet neutrinos should be dominated by the lightest m-doublet neutrinos.

For the other mass in \eqref{massneut}, without the information about the matrix $G$'s, we can only guess base on some assumption.  Specifically we will consider the possibility to have one of the lightest mass in $\mathcal{M}_3$ to be in the keV order.    First, it should be natural to assume $G_\nu$ and $G_e$ to have similar pattern (that is the mass pattern of o-doublet charged leptons).  From the pattern of $G_e$, the mass order of $M_{\nu L},$ and $M_{\nu L}'$ should be around $10^{-3} - 1$ GeV, while for $M_e$, $M_{\nu R}$, and $M_{\nu R}'$ should be around $\xi^{-1}$ times  $10^{-3} - 1$ GeV. Second, for simplicity assume that all entries of $\widehat{M}_m$ are around the same order.  To produce the correct mass pattern for the o-doublet neutrinos, the mass order of $\widehat{M}_m$ should be around $10^{10} - 10^{11}$ GeV.    If we want $\mathcal{M}_3$ to contain a keV mass order, then the largest order of $\mathcal{M}_4$ (or $D_b$) should be around $\xi^{-2}$ GeV.  

Unlike in the o-doublet charged lepton, in general $\mathcal{M}_3$ may not be diagonal, so the three generations of $E$ will mixed.  Thus even if the decay rate of m-doublet tauon and muon into m-doublet electron and neutrinos are very small, the mixing between them will make the lightest $E$ as the dominant component at low energy.  The same scenario also happen for the $\mathcal{M}_4$, and the lightest $E'$ should be the dominant component at low energy.

\section{BBN and the Dark Matters}
The model has two scalars, i.e. $\phi_e$ and $\phi_E$ that can act as inflaton fields during inflation epoch.  Since these  two scalars have different VEV, they can undergo different reheating scheme in the m- and o-sectors, and the two sectors can end up with different reheating temperature. Detail about this will be given in a future paper.   It is necessary that the reheating temperature be above the mass of the lightest singlet neutrinos $\nu_R$ and $N_L$, whose decay will lead to leptogenesis in both sectors, otherwise the particle-antiparticle asymmetry produced will be diluted by reheating. There are two possibilities regarding the reheating temperature, each will lead to different dark matter scenario.  

In the first possibility, the reheating temperature is below the mass of $h_R$, thus the mixing between $h_L$ and $h_R$ in the scalar potential is not effective anymore to make the two sector comeback to thermal equilibrium.  The two sectors will evolve with different temperature and this gives a solution for the BBN constraint, i.e. the m-sector has lower temperature than the o-sector.

Leptogenesis in this case take place after $h_R$ gain its mass.  
Therefore, lepton number produced in the m-sector cannot get converted into baryon number through the sphaleron process in m-sector (that took place around the time when $h_R$ gain its mass).  Thus there is no asymmetry in the m-baryon and all m-baryon will annihilate away to become m-mesons, dominated with m-pions.  These m-pions will decay through $W_R$ and $Z_R$ into m-doublet leptons.  When the temperature become very low, the m-sector will be dominated by m-doublet electrons and the lightest m-doublet neutrinos.  In this scenario there is no cold dark matter, only warm and some small fraction of hot dark matter, and we will not consider this case further.  

In the second possibility, the reheating temperature is far above the mass of $h_R$, and after reheating the two sectors will comeback into thermal equilibrium due to the mixing between $h_L$ and $h_R$.  As the temperature of the universe decrease, the massive $N_L$ and $\nu_R$ will decay into lighter fermions and will be the source of Leptogenesis mechanism, producing the same amount of lepton asymmetry in the m- and o-sectors.  This lepton (m-lepton) asymmetry will be converted into baryon (m-baryon) asymmetry through the sphaleron processes close to the electroweak symmetry breaking epoch in each sector, and the two sectors can have the same baryon asymmetry.  But there should be other process that make the two sectors to have different temperature before the BBN epoch.

The decay of the lightest $E'$ (or approximately the m-singlet electron) will provide large entropy contribution to the o-sector than to the m-sector.   The lightest $E'$ will decoupled from thermal equilibrium in the m-sector when its interaction rate $\Gamma < H$, the rate of cosmological expansion.   In the case when $\eta <<1$ the $E'$ has a very small m-electromagnetic charge, so its interaction with other m-fermion via m-photon is very small.  The same also for its interaction via $Z_R$, due to $Z_R$ being very massive and the $E'$ coupling via $Z_R$ is very small.  Via $h_R$ the $E'$ can interact with the m-doublet $E$ and other m-fermions.  But the coupling of $E'$ and $E$ with $h_R$, i.e. $G_e$ for the case of the lightest $E'$ is very small (the same Yukawa coupling of o-electron).  The $E'$ can also interact with o-fermions via $h_L$ due to its mixing with the singlet neutrinos $N'$ and $\nu'$, but with a very small mixing.  Therefore the lightest $E'$ will decouple from both sectors very early long before its decay, and thus after decoupling it can dominate the cosmic energy density.

The lightest $E'$ can decay into m-doublet electron $E$ and lighter m-fermions via $h_R$.  But, due to its mixing with $\nu'$ and $N'$, the lightest $E'$ can also decay with a larger rate into o-doublet neutrinos $\nu$ and lighter o-fermions via $h_L$.  With the same mixing the lightest $E'$ can also decay into m-doublet neutrino $N$ and lighter m-fermions via $h_R$, but with a smaller rate.  The ratio between the total decay rates into m- and o-fermions will determine the ratio of entropy contribution to the m- and o-sectors.  The total decay rate of $E'$ into o-fermions is given approximately by
\begin{equation}
\Gamma_{o} \approx \left(\frac{m_{E'}^5}{m_{h_L}^4}\right) \left(|\mathcal{U}_{E',5}G_\nu|^2 + |\mathcal{U}_{E',6}G_\nu'|^2\right)\frac{\sum_f |G_f|^2}{12 (8\pi)^3}
\end{equation}
where $G_f$ is the Yukawa coupling of the fermions and the sum is over o-fermions that have mass below the mass of $E'$, $m_{h_L}$ is the mass of $h_L$, and $U_{E',5/6}$ is the relevant entries of $M_E\widehat{M}_m^{-1}$ or $M_E'\widehat{M}_m^{-1}$.  While the total decay rate of the lightest $E'$ into m-fermions is given approximately by
\begin{eqnarray}
\Gamma_{m} \approx &&\left(\frac{m_{E'}^5}{m_{h_R}^4}\right) \left( |G_{e1}|^2+|\mathcal{U}_{E',5}G_\nu'|^2 + |\mathcal{U}_{E',6}G_\nu|^2\right)\nonumber \\ &&\times \frac{\sum_{f'} |G_f|^2}{12 (8\pi)^3}
\end{eqnarray}
where the sum is over m-fermions that have mass below the mass of $E'$, and $m_{h_R}$ is the mass of $h_R$.  The $G_{e1}\approx 3 \times 10^{-6}$ is the element of $G_e$ for the first generation (Yukawa coupling of the electron).  

Following the method in \cite{Kolb} one can get the ratio of the final entropy (after most off $E'$ have decayed) to the initial entropy per comoving volume in a particular sector 
\begin{equation}\label{Sformal}
\frac{S_f}{S_i} = \left[1+\frac{4}{3}\Gamma \left(\frac{45}{2\pi^2 g_*(T_i)} \right)^{1/3}\frac{m_{E'} Y_i}{T_i} I  \right]^{3/4}
\end{equation}
where $T_i$ is an initial temperature before the decay of the lightest $E'$, $\Gamma$ is the total decay rates of $E'$ into that particular sector, $Y_i = n_i R_i^3/S_i$, $n_i$ is the initial density of $E'$ after decoupling, $R_i$ is the cosmological scale at this initial time, $g_*$ is the relativistic degree of freedom in that particular sector during the decay process, and $I$ is some integral that contain a factor $<g_*>^{1/3}$, the average value of $g_*$ in that particular sector during the decay process.  We can assume that $g_*$ does not change appreciably in the two sectors.   If the time life of $E'$ is quite long the value of $I$ is large and the second terms inside the bracket of \eqref{Sformal} much larger than one, therefore the ratio of the final temperature between the m- and o-sectors is given approximately by
\begin{equation}
\frac{T_f'}{T_f}=\left(\frac{S_f'}{S_f}\right)^{1/3} \approx \left(\frac{\Gamma_m}{\Gamma_o} \right)^{1/4}.\nonumber
\end{equation} 
Assuming $m_{h_L}/m_{h_R}\approx \xi$ we have
\begin{equation}
\frac{T_f'}{T_f}\approx r^{1/4} \xi  \left(1+\frac{|G_{e1}|^2}{|\mathcal{U}_{E',5}G_\nu'|^2+|\mathcal{U}_{E',6}G_\nu|^2} \right)^{1/4},
\end{equation} 
where $r = \sum_{f'} |G_f|^2/\sum_{f} |G_f|^2 < 1$.  Since $G_{e1}$ is very small, it is possible that $|\mathcal{U}_{E',5}G_\nu'|^2,|\mathcal{U}_{E',6}G_\nu|^2>> |G_{e1}|^2$, in which case $T_f'/T_f\approx r^{1/4}\xi$.   

In the m-hadronic sector, the m-baryon-antibaryon will annihilate through SU(3)$_2$ interaction to become m-mesons, dominated by m-pions, leaving the asymmetric part of the m-baryon that will decay through the m-weak gauge bosons $W_R$ into lighter m-baryons. While the m-pions will decay through the m-weak interaction to become m-leptons.  In the end, in the m-hadronic sector we are left with the asymmetric part of m-nucleons, i.e. the m-proton and m-neutron.  These m-nucleons can form a m-nucleus, a collection of m-nucleon bound together with m-nuclear force whose strength is similar to the o-nuclear force.  Since the m-electromagnetic force repulsion between m-protons are very weak, the m-nucleus can be very large.  This large m-nucleus form the cold dark matter component that have some self-interaction with the strength similar to o-nuclear force.  Assuming the mass of m-proton and m-neutron are approximately the same, then the m-nucleon will contribute energy density $\rho_b' = n_b m_{p'}$, where $n_b$ is the o-baryon density and $m_{p'}$ is the mass of m-proton.  If all m-nucleons are the dark matter, then $m_{p'}$ should be around five times the mass of o-nucleon, but if there are other significant components of dark matters then $m_{p'}$ should be less than five time proton mass.        

The mass of o- or m-nucleon depends on the hadronic scale in that sector, i.e. $\Lambda_{QCD}$ and $\Lambda_{QCD}'$ respectively, and on the mass of their quark constituent.  We can write for the case of o-proton and m-proton respectively \cite{anh}
\begin{eqnarray}\label{massmproton}
m_{p} &=& k\Lambda_{QCD}+2 h_{pu} m_{u} + h_{pd} m_{d},\nonumber\\
m_{p'} &=& k'\Lambda_{QCD}'+2 h_{pu}' m_{u'} + h_{pd}' m_{d'},
\end{eqnarray}
where $k^{(')},h_{pu}^{(')}, h_{pd}^{(')}$ are some parameters, $m_{u^{(')}}$ and $m_{d^{(')}}$ are the mass of o- (m-) up and o- (m-) down quarks respectively.  Similar formula can be written also for the o- and m-neutron.  The mass of a m-quark is $\xi^{-1}$ times the mass of the corresponding o-quark.  The hadronic scale in the m-sector is related to the hadronic scale in the o-sector \cite{anh}
\begin{eqnarray}\label{lambdaqcd}
\Lambda_{QCD}' &=& \left(m_u m_d m_s\frac{\Lambda_{QCD}^{9}}{\xi^4}\right)^{1/11},\  \text{for}\ \Lambda_{QCD}' < m_{u'},m_{d'}\nonumber\\
\Lambda_{QCD}' &=& \left(m_d m_s\frac{\Lambda_{QCD}^{27/2}}{\xi^5}\right)^{2/31}, \  \text{for}\ m_{u'} <\Lambda_{QCD}' < m_{d'}\nonumber\\
\Lambda_{QCD}' &=& \left(m_s \frac{\Lambda_{QCD}^{27/2}}{\xi^4} \right)^{2/29}, \  \text{for}\  m_{u'},m_{d'}< \Lambda_{QCD}'< m_{s'}\nonumber\\
\end{eqnarray}
where $m_s$ are the mass of o-strange quark.  The three relation above is actually approximately the same.   

If the mass of the quark constituent is very low compared to the hadronic scale, then the mass of the nucleon is proportional to the hadronic scale.   While if the quark constituent is heavier than the hadronic scale, then they will behave non relativistically.  In this later case the mass of m-proton will be the mass of its quarks constituent plus a negative small contribution from the gluonic interaction.   We can assume that the parameter $k^{(')}$ in \eqref{massmproton} is changing as a function of the quark constituent mass. When the mass of the quark constituent approaching $\Lambda$, the value of $k'$ should decrease, diminish then negative. While when the mass of the quark constituent is smaller compared to $\Lambda$ the value of $k^{(')}$ will approach a maximum value $k_{max}$.

Since in the o-sector the mass of the quark constituent is very small compared to the $\Lambda$, we can assume the parameter $k$ there is close to $k_{max}$.  Lets take $m_u \approx 2.5$ MeV, $m_d \approx 5$ MeV, $m_s\approx 95$ MeV, and $\Lambda_{QCD}\approx 200$ MeV, then using $m_p =938$ MeV, we have $k_{max}\approx 4.7$.   For $\xi \approx 0.1$ we have to take the case when $ m_{u'},m_{d'}< \Lambda_{QCD}'< m_{s'}$ in \eqref{lambdaqcd}, and thus $\Lambda_{QCD}' \approx 359 $ MeV.  With this value, both the m-down and m-up quark should behave relativistically.  Therefore we can set $h_{pu}',h_{pd}'\approx 1$, while $k'$ will not far below  $k_{max}$.  The mass of m-proton is then
\begin{equation}
m_{p'}\approx k_{max} \Lambda_{QCD}'+\xi^{-1}(2 m_u + m_d).\approx 1.8 \text{ GeV}. 
\end{equation}
The correct value of $k'$ can be found using Lattice QCD, but we can conclude that for $\xi \approx 0.1$, the contribution of m-nucleon for the dark matter density is still less than $5\Omega_{B}$.   The other contribution for $\Omega_{DM}$ should then come from the m-lepton sector.

In the m-lepton sector, after the decay of the m-singlet electron what is left are the m-doublet leptons.  The m-doublet tauon and muon will decay through $W_R$ to become m-doublet neutrinos and the m-doublet electrons.  These last particles are in thermal equilibrium due to m-weak interaction and m-electromagnetic interaction.  Since the m-photon mass is less than the mass of the m-weak gauge bosons $W_R$ and $Z_R$, the m-doublet electron will decoupled from the m-doublet neutrinos when the rate of  m-electromagnetic interaction between them are less than the cosmological expansion rate $H$.  Since $m_D$ is large, the m-electromagnetic interaction rate between m-doublet electron and m-doublet neutrinos is very small. For example in the relativistic regime, the interaction rate of $e'+\nu' \rightarrow e'+\nu'$ is given by 
\begin{equation}
\Gamma_{e'\nu'} = \frac{q_e^2q_\nu^2 \mathcal{E}^2}{\pi m_D^4}\frac{\xi(3)}{\pi^2}T^3
\end{equation}
where $-q_e, q_\nu \approx g/2$ are the fractional m-electromagnetic charge of m-doublet electron and m-doublet neutrino, and $\mathcal{E}$ is the energy of the particle (in the center of momentum frame).  It is important that the m-doublet electron decouple from the cosmic plasma in the m-sector after the decay of $E'$, otherwise its energy density will over-close the universe.   Taking $m_D = 800$ TeV, the decouple temperature depend on the ratio $x\equiv T'/T$.  If $x=0.1$ then the decoupling temperature $T_d \approx 2.2$ GeV.  Thus the keV m-doublet electrons will decouple from the m-doublet neutrinos when it is still relativistic and it will become the warm dark matter component.   

The m-doublet electron energy density today is given by $\rho = n_E m_E (R/R_0)^3$, where $n_E$ is its number density when at the decoupled time, $m_E$ is its mass, $R_0$ and $R$ is the present day and at the decoupled time length scales respectively.    Assuming that the entropy density in the m-sector today is very small, we can use the present day photon entropy density $s_0 = 2970$ cm$^{-3}$ and $s$ the entropy density at the time of m-doublet electron decoupled from the cosmic plasma, to give the length scale change of the universe $R/R_0 = (s_0/s)^{1/3}$.  The relative energy density of the m-doublet electron is given by
\begin{equation}
\Omega_E\approx\frac{n_E R^3}{\rho_c R_0^3}m_E = \frac{40.57}{g_s^*\pi^4}x^3 \frac{2970}{0.52 \times 10^{-5}}m_E(\text{GeV}) 
\end{equation}
where $g_s^*$ is the o-sector entropy relativistic degree of freedom at the decouple time of $E$.  For $m_E = 7.1$ keV, $x=0.1$ and $T_d\approx 2.2$ GeV we have $g_s^* = 72.25$, and $\Omega_E \approx 0.5\ \Omega_B$.  While if $x=0.2$ and $T_d\approx 0.7$ GeV, we have $g_s^* = 61.75$, and $\Omega_E \approx 4.8\ \Omega_B$.  If $x$ is very small, then the model will be dominated by a cold dark matter with very small warm dark matter contribution to $\Omega$.   

\section{Conclusion} 
The introduction of new scalars $\phi_e$ and $\phi_E$ to the mirror model turns out to give many new phenomena that has been elaborated above.  Even though many of the result depend on unknown parameters value, but reasonable assumption on those parameters shows that this modified mirror model can become a good candidate for mixed cold-warm dark matter scenario.  The ratio of the cold-warm dark matter component depend indirectly to the ratio between the VEVs, i.e. $\xi$.  So once a more accurate value of $\xi$ is known, either from the mixing of the SM-Higgs with other heavier scalar (the m-Higgs) or from the sterile-active neutrinos mixing, many of the feature of this model can be tested.

\end{document}